\documentclass[twoside,a4paper]{msmb} % Класс документа [twoside,a4paper]
\begin{document}
\yearpub{2020}                                      % Год издания. Заполняет редакция.
\num{3}                                             % Номер журнала. Заполняет редакция.
\volume{55}                                         % Сквозная нумерация, том, выпуск
\setcounter{page}{12}
%%%%%%%%%%%%%%%%%%%%%%%%%%%%%%%%%%%%%%%%%%%%%%%%%%%%%%%%%%%%%
%%%%%%%%%%%  Общие праметры для вёрстки и оформления  %%%%%%%%%%%%%%%%%
%%%%%%%%%%%%%%%%%%%%%%%%%%%%%%%%%%%%%%%%%%%%%%%%%%%%%%%%%%%%%
\renewcommand{\filename}{scepl}                  % Вставьте имя вашего файла.
%\startarticle
\date{05.10.2020}
\shrtitle{Частицы-призраки, сцепленность исторических эпох...}         % Часть названия статьи для колонтитулов.
\shrauthor{А.К.~Гуц}{А.К.~Гуц}  % Автор или группа авторов для оглавления и колонтитулов.
\udc{530.145}                               % Вставьте УДК.
\selectlanguage{english}                    % Основной язык статьи
\date{05.10.2020}
%%%%%%%%%%%%%%%%%%%%%%%%%%%%%%%%%%%%%%%%%%%%%%%%%%%%%%%%%%%%%%%%%%%%%%%
%%%%%%%%%%%% для русскоязычной статьи и реферирования в РИНЦ  %%%%%%%%%
%%%%%%%%%%%%%%%%%%%%%%%%%%%%%%%%%%%%%%%%%%%%%%%%%%%%%%%%%%%%%%%%%%%%%%%
\title{Частицы-призраки, сцепленность исторических эпох и~машина времени}
\author{А.К.~Гуц}{д.ф.-м.н., профессор}{guts@omsu.ru}
\affil{Омский государственный университет им.~Ф.М.~Достоевского, Омск, Россия}
\abstract{В статье изучается возможность создания машины времени,
основанной на механизме квантового сцепление макроскопических
обычных (много)частичных конфигураций и (много)частично-призрачных
конфигураций различных исторических эпох, принадлежащих различным
параллельным эверетовским вселенным. }

\titleEng{Ghost Particles, Entanglement of~Historical Epochs and~Time Machine}
\authorEng{A.K.~Guts}{Dr.Sc. (Phys.-Math.), Professor}{aguts@mail.ru}
\affilEng{Dostoevsky Omsk State University, Omsk, Russia}
\abstractEng{In the article the possibility of creating a time
machine, based on the mechanism of quantum entanglement of
macroscopic ordinary (many)partial configurations and
(many)partially ghostly configurations of different historical
epochs belonging to different parallel Everett universes is
investigated.} \keywordsEng{Time machine, ghost particles,
entanglement historical epochs, parallel universes}
%%%%%%%%%%%%%%%%%%%%%%%%%%%%%%%%%%%%%%%%%%%%%%%%%%%%%%%%%%%%%%%%%%%%%%%%%%%%%%%%

\maketitle
 % Формирование оглавления.

%\usepackage{floatflt}

%%%%%%%%%%%%%%%%%%%%%% Т Е К С Т %%%%%%%%%%%%%%%%%%%%%%%%

\def\R{{\mathbb{R}}}

\section*{Introduction}
Deutsch in <<The Fabric of Reality>> \cite{doj} universe, that is
parallel to our universe in the sense of the Everett's
interpretations of quantum mechanics, called {\it shadow}.

Elena Palesheva in the article \cite{pal} linked shadow particles
of parallel universe with {\it ghost particles} of our Universe.
She also confirmed the Deutsch's point that shady particles, i.e.
ghost particles can weakly interact with ordinary particles of our
universe through quantum interference.

In the \cite{aguc,aguc1,aguc16} articles we suggested linking
space-time trajectories appearing in Wheeler-DeWitt
geometrodynamics with real-life parallel historical epochs, which
are various time periods of human civilization. Transition from
one epoch to another carried out by launching a special apparatus,
called a time machine, was realized as the mechanism of quantum
entanglement of compact spatio-temporal regions of different
historical epochs. However, in this article we did not say how
this entanglement
 occurs.

In given article we proposes to consider as such mechanism the
entangling of macroscopic usual  (many) partial configurations and
(many) partial-ghost configurations. The latest are the
configurations from the parallel universe due to the idea of Elena
Palesheva.

\section{Ghost particles}

A ghost particle is a particle whose moment-momentum tensor is
zero, but its current is non-zero, which in in the case of a
bispinor is
$$
j^i=\psi^+\gamma^i\psi.
$$
Therefore, such a particle carries neither energy nor momentum.

Such non-Abelian solutions of the Yang--Mills equations were found
by Loos in 1967 \cite{loos}. Recall that the quanta of the
Yang--Mills fields are vector particles (i.e. bosons with spin 1)
and have zero mass. However, through the mechanism of spontaneous
symmetry breaking physical Yang--Mills fields can acquire non-zero
mass. Under considering the weak interaction  the quantum of
Yang--Mills fields is considered a W-particle having a charge of
+1, 0 or -1. For the strong interaction, the quantum of the
Yang--Mills field is gluons hold protons and neutrons together.

 In the case of particles propagating in outer space, ghost particles
were discovered in 1974 by Griffith \cite{grif}, and then the
corresponding solutions of the Einstein--Dirac equations were
found and published in works
\cite{dav,davis2,let,nov,aud,devis,grif2,guc,mih,grif3,der,dem,cast0,cast}
1970-80s of the XX century, and already in the XXI century ---
\cite{gar,han,davis3,her,wil,tal,zek}. Neutrino ghosts and massive
bisprinor ghosts were found in the universes Robinson-Walker, i.e.
in Friedmann's universes \cite{zek}. "What the most interesting
thing about ghost solutions is how the neutrino field propagates
in the background space-time without changing him, which means we
can't detect ghost neutrinos by their gravitational effects"
\cite{han}.

"The first reaction, writes M. Novello in the article "The ghostly
foundations for neutrinos"\, (1974), --- these solutions may be
rejected on physical grounds. However, the fact
  that they seem to be present in any geometry\footnote{It is stated in \cite{aud} that
   ghost neutrinos (described by the general covariant Weyl equation $
\varphi_{||A\dot{X}}^{A}=0$) exist only in algebraic special
space-times
    with a neutrino flux vector parallel to one of the main zero
    vectors of the conformal tensor.},
  makes their less superficial study expedient.
  To be able to study some of the details of these ghosts,
  we must find not only one, but also a class of these solutions
   in the given geometry".
And he discovers a surprising result: he gives an example
space-time generated by a neutrino, and this itself neutrino is a
linear combination of neutrino-ghosts \cite{nov}.

\medskip

However, before the work of Elena Palesheva no one gave at least
some interpretations to ghost particles.

Combining the results of Palesheva and Novello, we can state the
following hypothesis.

The universe, in which we are aware of our presence, consists of
real particles, i.e. particles with a non-zero tensor
energy-momentum. Ghost particles are guests from parallel
universes. But there are infinitely many parallel universes; all
of them are symmetrical with respect to our analysis (there is no
dedicated "our"\ Universe), therefore, one can only exist ghost
particles. Energy and momentum are imparted to a particle from the
specific one under consideration, i.e., fixed by someone
consciousness, the universe, if, from the point of view of
mathematics, it is linear combination of ghost particles. But to
decompose a particle into linear combination requires some
mechanism present in universe, which brings about and confirms the
fact of decay.

Obviously, this is the same mechanism that fixed a particular
universe. And this mechanism is consciousness, there is an
observer, present, living in this universe.

Consciousness, by its attention to the universe surrounding it,
performs an act creation, expressed in the generation of linear
combinations ghost particles: from chaos of ghost particles to
order="linear combinations", from simple to complex.

It is important that Palesheva showed that ghost particles, i.e.
particles parallel universe, interact with our particles and this
interaction manifests itself in the form of quantum interference.

However, another interaction of the particles of our Universe is
possible and parallel. This is quantum entanglement
(entanglement). Let's describe it the {\it non-force} interaction
of particles from different universes.

\section{Examples of ghost particles}

  Consider spinor particles described by the equation
Dirac:
\begin{equation}\label{d}
i\hbar {\gamma}^{(k)}\frac{\partial\psi}{\partial
x^{\scriptscriptstyle k}}-mc\psi =0.
\end{equation}
Then bispinor
\begin{equation}\label{sp}
\psi(x)=\left[\begin{array}{r} 1\\ 1\\ -1\\
1\end{array}\right]e^{\frac{mc}{\hbar}x^{\scriptscriptstyle
2}+f(x^{\scriptscriptstyle 0}+x^{\scriptscriptstyle
3})+ig(x^{\scriptscriptstyle 0}+x^{\scriptscriptstyle 3})}
\end{equation}
is a solution to the Dirac equation. Here $g(x^{\scriptscriptstyle
0}+x^{\scriptscriptstyle 3})$ and $f(x^{\scriptstyle 0}
+x^{\scriptscriptstyle 3})$ --- smooth real functions.

Based on the results of Theorem 12.1 from \cite{guc1,Pal2}, we
obtain that (\ref{sp}) describes a spinor ghost only if
$g(x^{\scriptscriptstyle 0}+x^{\scriptscriptstyle 3})=const\in
\R$.

Let's take the solution for a real wave in the form:
\begin{equation}\label{sp1}
\psi(x)=\left[\begin{array}{r} 1\\ 1\\ -1\\
1\end{array}\right]e^{\frac{mc}{\hbar}x^{\scriptscriptstyle
2}+i(x^{\scriptscriptstyle 0}+x^{\scriptscriptstyle 3})},
\end{equation}
and for the spinor ghost we set
\begin{equation}\label{sp2}
\phi(y)=\left[\begin{array}{r} 1\\ 1\\ -1\\
1\end{array}\right]e^{\frac{mc}{\hbar}y^{\scriptscriptstyle 2}}.
\end{equation}

\section{Particle entanglement and ghost particles}

Let $|\psi(\pm)\rangle_r$ and $|\phi(\pm)\rangle_g$ be the states
of a real particle and a ghost corresponding to bispinor solutions
$\psi(x,\pm)$ and $\phi(x,\pm))$ of Dirac equations with different
spin projections $\pm$ \cite{Bit}.

The entangled state of two particles in this case is described by
ket-vector
$$
|\psi(+)\rangle_r \otimes|\phi(-)\rangle_g+|\psi(-)\rangle_r\
\otimes|\phi(+)\rangle_g.
$$
According to the \textbf{ER=EPR} statement about the existence of
an Einstein-Rosen bridge, or a 3-dimensional wormhole, connecting
the locations of entangled  particles (EPR-pairs) \cite{Mald}, we
can say that there is a 3-dimensional wormhole connecting a
particle of our universe with a shadow particle, or ghost
particle, from a parallel universe.

\section{How to entangle the particles?}

"How to entangle the particles: take a crystal with non-linear
optical properties --- that is, such, the interaction of light
with which depends on the intensity of this light. For example,
triborate lithium, barium beta-borate, potassium niobate.
Irradiate it with a laser appropriate wavelength --- and
high-energy laser photons radiation will sometimes decay into
pairs of entangled photons less energy (this phenomenon is called
"spontaneous parametric scattering") and polarized in
perpendicular planes. It remains to keep the entangled particles
intact and spread as far apart as possible" \cite{vas}.

As you can see, the entanglement mechanisms are already in place.
Although not yet those we need.

\section{Macroscopic entanglement}

To achieve our goal, we need a macroscopic multiparticle
entanglement. Does it exist? Common the opinion that macroscopic
entanglement does not occur in nature outside artificially created
situations. What is this opinion based on?

It is known that as the size of physical systems increases it gets
{\it harder completely
  isolate them from the environment}, and interaction with
  the environment --- decoherence ---
  destroys quantum characteristics such as superpositions
and entanglement, i.e., in particular, the entanglement
many-particle configurations from parallel epochs.

The beauty of a quantum description of an object is that such an
object  is a quantum object represented by a coherent
superposition waves, which has both an interference pattern and,
possibly, entanglement. But even if there is macroentanglement, it
is difficult to discover   on a macroscopic scale, since quantum
objects interact with a large number of degrees of freedom, and
any inaccuracy in the original data or follow-up (coarse-grained)
leads to decoherence, loss of consistency between the waves
describing all these degrees of freedom, and to the disappearance
of their interference pattern, to the loss of entanglement.
Consequently, the essence of the quantum  descriptions,   giving
fantastic from the point of view of the classical physics
possibilities is lost.

For our goal -- the transfer of a macro-object, a person to
another epoch --- it is necessary to make sure that the laws of
quantum mechanics are also valid for macroobjects. To do this, you
need to pay attention to those of their mechanical degrees of
freedom of the macro object, which are well isolated from the
environment and therefore well protected from decoherence. We need
macroscopic quantum mechanics.

 The appearance of such a "mechanics"\ became
possible thanks to the recent progress in quantum optomechanics:
physicists were able to use light (sometimes microwaves) to
transform macroscopic mechanical objects into almost pure quantum
states <..>. Soon they will be able to allow these mechanical
objects to evolve without much decoherence and measure the final
states, thereby making comparisons with the predictions of quantum
mechanics" \cite{chen}.

Note that we are using the terms "macroscopic"\ on a purely
intuitive level, and it's not very correctly with a scientific
approach to the problem under study.

It is important to decide in advance
  with what is considered a macroscopic object:
"although quantum systems are difficult to maintain and observe in
macroscale, they can be easily created. On the other hand, the
question naturally arises: is the received macroscopic state?
Based on experimental results, we have shown that an entangled
state that can be be obtained with the help of currently available
technologies, will include a sufficiently large number of photons,
which can be seen with the naked eye" \cite{sek0}. This makes our
approach is satisfactory if macroscopicity is concept related to
size. We also mentioned that the components entangled state can be
easily distinguished by a simple avalanche photodiode, if you look
at the dispersion of the distribution of the number photons. This
pleases those who believe that macro-entangled states should have
components that can be easily distinguished. Although our study
showed that the resulting state was surprisingly resilient to
losses, we showed that it also becomes more and more brittle under
phase perturbation at increasing its size. So our approach is also
satisfactory if macroscopic agents are sensitive to decoherence,
and highlights the complexity of possible interactions between a
given quantum system and its environment. We also saw that the
accuracy of measurement needed to detect the quantum nature of the
created state, increases with its size. This also makes our scheme
satisfactory if macroscopicity associated with the requirement for
measurement accuracy. In conclusion, we note that there are many
other candidates for a measure of macroscopicity \cite{fro}.
Testing each one is "working for the future" \cite{sec0}.

\section{Entanglement and wormholes}

Suppose that in the near future we will learn how to create
macroscopic entangled multiparticle configurations. But for our
goals related to the time machine, this is not enough, it is
necessary chain configurations with ghost configurations.

What should we expect if we manage to do this?

The quantum phenomenon of entanglement is closely related to the
classical the phenomenon of the formation of a 3-dimensional
wormhole. Consequently, entanglement in space will give rise to a
3-dimensional wormhole or 4-dimensional wormhole between parallel
universes, between various historical epoches. Transitions throut
such a wormhole are the quantum time machine \cite{aguc,aguc1}.

However, since 3-dimensional wormholes are unstable, it follows
think about generating 4-dimensional wormholes \cite{akguc}. Last
thing, most likely indicates entanglement in time \cite{now} (Time
Entanglement), i.e. the formula \textbf{EPR=ER} is replaced by the
formula \textbf{EPR=TE}.

{\protect\footnotesize \selectlanguage{english}

}
%\newpage

\endarticle


\begin{thebibliography}{99}
\bibitem{doj}
Deutsch D. The Fabric of Reality.  Penguin Books, London, 1997.


\bibitem{pal}
Palesheva E.V. Ghost spinors, shadow electrons and the Deutsch
Multiverse.  \url{arXiv:gr-qc/0108017v2} (2001).


\bibitem{aguc}
Guts A.K. Vremennye effekty kollapsa volnovogo paketa v
superprostranstve Uilera. Mezhdunarodnyi nauchnyi seminar
<<Nelineinye modeli v mekhanike, statistike, teorii polya i
kosmologii>> GRACOS-16, Lektsii shkoly i materialy seminara (5--7
noyabrya 2016 g., Kazan'). Kazan', Kazanskii (Privolzhskii)
federal'nyi universitet, 2016, pp.~273--280.
 (in Russian)

\bibitem{aguc1}
Guts A.K.  Quantum time machine. Prostranstvo, vremya i
fundamental'nye vzaimodeistviya, 2019, no.~3, pp.~20-44.
 (in Russian)

\bibitem{aguc16}
Guts A.K. Geometry of historical epoch, the Alexandrov's problem
and non-G\"odel quantum time machine. e-Print archive: 1608.08532
(2016). URL: http://arxiv.org/abs/1608.08532v2

\bibitem{loos}
 Loos H.G. Free Ghost Gauge Fields. Nuovo cimento, 1967, vol.~LII~A, no.~4,
 pp.~1085--1091.


\bibitem{grif}
Griffiths J.B.  Gravitational radiation and neutrinos.
Communications in Mathematical Physics, 1972, vol.~28, pp.~295--299.

\bibitem{davis2}
Davis T.M. and Ray J.R.  Ghost neutrinos in general relativity.
Physical Review D., 1974, vol.~9, iss.~2, pp.~331--333


\bibitem{dav}
Davis T.M. and Ray J.R. Ghost neutrinos in plane-symmetric spacetimes.
Journal of
 Mathematical Physics, 1975, vol.~16, pp.~75.


\bibitem{natur}
Ghost neutrinos emerge from mathematics. Nature, 1974, vol.~248,
pp.~471--472.


\bibitem{let}
Letelier P.S. Ghost neutrons in the Einstein-Cartan theory of
gravitation.
Physics Letters A., 1975, vol.~54, issue~5,  pp.~351-352


\bibitem{nov}
Novello M. Ghost basis for neutrino. Physics Letters A., 1976,
vol.~58, iss.~2, pp.~75--76.
 (in Russian)

\bibitem{aud}
Audretsch J. Ghost neutrinos as test fields in curved space-time l.
Physics Letters A., 1976, vol.~56, iss.~1, pp.~15-16.



\bibitem{devis}
Davis T.M. and Ray J.R. Neutrinos and Bianchi I universes. Journal
of Mathematical Physics 17, 1049 (1976).



\bibitem{grif2}
 Griffiths J.B. On the propagation of photons and neutrinos in
curved space-time. General Relativity and Gravitation, 1977, vol.~8,
pp.~365--370.



\bibitem{guc}
Guts A.K. A new solution of the Einstein-Dirac equations.
Izvestiya vuzov. Fizika, 1979, no.~8. pp.~91--95.
 (in Russian)

\bibitem{mih}
Michalik T.R. and   Melvin M.A. Spatially homogeneous neutrino
cosmologies. Journal of Mathematical Physics, 1980, vol.~21,
pp.~1952--1964.


\bibitem{grif3}
Griffiths J.B. Ghost neutrinos in Einstein–Cartan theory. Phys.
Lett., 1980, A75, pp.~441-–442.


\bibitem{der}
Dereli T. and Tucker R.W. Exact neutrino solutions in the presence of
torsion. Physics Letters A., 1981, vol.~82, iss.~5, pp.~229--231.


\bibitem{dem}
Demakis A. and M$\ddot{u}$ller-Hoissena F. Massive ghost Dirac fields
in Einstein-Cartan theory. Physics Letters A., 1982, vol.~92,
iss.~9, pp.~431--432.


\bibitem{dem1}
Demakis A. and  M$\ddot{u}$ller-Hoissena F.   Solutions
of the Elnsteln-Cartan-Dirac equations with vanishing
energy-momentum tensor. Journal of Mathematical Physics, 1985,
vol.~26, pp.~1040--1048.



\bibitem{cast0}
Torres del Castillo G.F.. Wavelike solutions to the Einstein
equations coupled to neutrino and gauge fields. Journal of
Mathematical Physics, 1986, vol.~27, pp.~2756.

\bibitem{cast}
Torres del Castillo G.F.   Ghost Neutrino Fields in Flat
Space-Time. General Relativity and Gravitation, 1987, vol.~19,
no.~7,  pp.~699--705.


\bibitem{gar}
Garcia de Andrade L.C.. Ghost neutrinos and radiative Kerr metric
in Einstein-Cartan gravity. \url{arXiv:gr-qc/0204084v1} (2002).

\bibitem{han}
Muxin Han, Yapeng Hu, and Hongbao Zhang. Exhaustive Ghost Solutions to
Einstein-Weyl Equations for Two Dimensional
 Spacetimes. 2004. URL: \url{http://arxiv.org/abs/gr-qc/0409019v2}.


\bibitem{davis3}
 Davis T.M. and Ray  J.R. Neutrinos and Bianchi I universes.
Journal of Mathematical Physics, 1976, vol.~17, pp.~1049.


\bibitem{her}
  Herrera L. and Jimenez J. Neutrino fields in axially and reflection symmetric
space-times. Journal of Mathematical Physics, 1979, vol.~20(1),
pp.~195--198.



\bibitem{wil}
Wils P. A class of exact solutions of the Einstein–Dirac equations.
Journal of Mathematical Physics, 1991, vol.~32, pp.~231-233.




\bibitem{tal}
 Talebaoui W. Non-ghost massless solution of the
Einstein-Dirac field equations. Class. Quantum Grav, 1995,
vol.~12,  pp.~2051-2057.




\bibitem{zek}
Zekka A. The Einstein-Dirac Equation in Robertson-Walker
Space-Time Does Not Admit Standard Solutions. Int. J. Theor.
Phys., 2009, vol.~48, pp.~2305-2310.


\bibitem{guc1}
Guts A.K. The Elements of Time theory. Moscow, URSS Publ.,~2012.
 (in Russian)


\bibitem{Pal2}
Palesheva E.V. Contribution of spinor ghosts to interference of
quantum particles // Mathematical structures and modeling. 2002.
№~9. P.~142--157.


\bibitem{Bit}
 Bittencourt S.V., Bernardini A.E., and Blasone M.
 Efects of Lorentz boosts on Dirac bispinor
entanglement. \url{arXiv:1810.01568v1} (2018).


\bibitem{Mald}
Maldacena J. and Susskind L. Cool horizons for entangled black holes.
\url{arXiv:1306.0533}


\bibitem{vas}
Vasil'ev S. Kvantovaya zaputannost' -- koroleva paradoksov. URL:
\url{https://naked-science.ru/article/nakedscience/kvantovaya-zaputannost}.
(in Russian)


\bibitem{chen}
 Chen Y. Macroscopic quantum mechanics: theory and experimental
concepts of optomechanics. Journal of physics B: atomic,
molecular and optical physics, 2013, vol.~46, pp.~104001.


\bibitem{sek0}
Sekatski P., Sanguinetti B. et al. Cloning Entangled Qubits to
Scales One Can See. URL: \url{https://arxiv.org/abs/1005.5083v1} (2010).




\bibitem{fro}
Frowis F. and D\"ur W. Measures of macroscopicity for quantum spin
systems. New Journal of Physics, 2012, vol.~14, pp.~093039--093062.



\bibitem{sec0}
Sekatski P., Sangouard N., Stobinska M.,  Bussieres F.,
 Afzelius M., and  Gisin N. Proposal for Exploring Macroscopic Entanglement with
a Single Photon and Coherent States. 2013. URL: \url{http://arXiv:1206.1870v2}
[quant-ph].


\bibitem{sek1}
Sekatski P., Sangouard N., and Gisin N.
 Size of quantum
superpositions as measured with classical detectors.  Phys.
Rev. A., 2014, vol.~89, pp.~012116.


\bibitem{mart}
De Martini F. Entanglement and Quantum Superposition of a
Macroscopic Macroscopic system. 2010.
URL:\url{https://arxiv.org/abs/0903.1992v2}.


\bibitem{kitaev}
Kitaev A. and Preskill J. Topological entanglement entropy. 2006. URL:
\url{https://arxiv.org/abs/hep-th/0510092v2}.

\bibitem{akguc}
Guts A.K. Physics of Reality. Omsk, izd-vo KAN, 2012.
 (in Russian)

\bibitem{now}
Nowakowski M. Quantum Entanglement in Time. AIP Conference
Proceedings, 2017, vol.~1841, pp.~020007. URL: \url{https://doi.org/10.1063/1.4982771}.
(in Russian)

\end{thebibliography}
\end{document}